\documentstyle[]{article}
\def\1ad{\mbox{\normalsize $^1$}}
\def\2ad{\mbox{\normalsize $^2$}}
\def\3ad{\mbox{\normalsize $^3$}}
\def\4ad{\mbox{\normalsize $^4$}}
\def\5ad{\mbox{\normalsize $^5$}}
\def\6ad{\mbox{\normalsize $^6$}}
\def\7ad{\mbox{\normalsize $^7$}}
\def\8ad{\mbox{\normalsize $^8$}}
\def\makefront{\vspace*{1cm}\begin{center}
\def\newtitleline{\\ \vskip 5pt}
{\Large\bf\titleline}\\
\vskip 0.5truecm
{\large\bf\authors}\\
\vskip 3truemm
\addresses
\end{center}
\vskip 0.5truecm
{\bf Abstract:}
\abstracttext
\vskip 0.5truecm}
\def\sn{\smallskip\noindent}

\def\pano{\par\noindent}

\def\IP{\relax{\rm I\kern-.18em P}}
\def\la{\langle}
\def\ra{\rangle}
\def\P{\,\hbox{\hbox to -0.2pt{\vrule height 6.5pt width .2pt\hss}\rm P}}

\def\ts{\textstyle}

\def\beq{\begin{equation}} \def\eeq{\end{equation}}
\def\bea{\begin{eqnarray}} \def\eea{\end{eqnarray}}
\def\bc{\bar c}   
  \def\bJ{\bar J}

\def\vchi{\vec \chi}

\def\tabroom{\hbox to0pt{\phantom{\Huge A}\hss}}
\def\tab2{\hbox to0pt{\phantom{\huge A}\hss}}
\def\btau{\bar \tau}  
\begin {document}
\rightline{\vbox{\hbox{hep--th/9901005}
 \hbox{HUB--EP--99/02}}}
\def\titleline{Issues of Heterotic (0,2) Compactifications}
\def\authors{Ralph Blumenhagen}
\def\addresses{
Humboldt Universit\"at Berlin\\
Institut f\"ur Physik, Invalidenstrasse 110, 10115 Berlin, Germany
}
\def\abstracttext{
A brief review of some aspects of  heterotic $(0,2)$ compactifications in the
framework of exactly solvable superconformal field theories and
gauged linear sigma models is presented. 
}
\large
\makefront

\section{Introduction}
The general class of four dimensional $N=1$ supersymmetric heterotic 
compactifications has $(0,2)$ supersymmetry on the world sheet. Geometrically, 
such models are defined by a stable holomorphic vector bundle (coherent
sheaf) over a Calabi-Yau threefold subject to further anomaly 
cancellation conditions 
involving the first and  second Chern classes of the vector and the 
tangent bundle. 
It has long been an open question whether generically such models
are indeed consistent vacua of the heterotic string. It has been argued
in \cite{paper1} that $(0,2)$ models realizable as linear sigma models
\cite{paper2}
provide a set of exact perturbative vacua. 
In \cite{paper3} we presented a class of exactly solvable superconformal
field theories with $(0,2)$ supersymmetry, which were argued to describe
special points in the moduli space of $(0,2)$ linear sigma models.
As a byproduct we got to know a subset  of $(0,2)$ models inheriting
their defining data from $(2,2)$ models and automatically satisfying all
the linear and quadratic anomaly constraints. 
As a first step towards establishing mirror symmetry in the $(0,2)$ context
we defined a way to obtain mirror symmetric pairs at least in the 
afore mentioned subclass
of models \cite{paper4}. Another  approach was made in \cite{paper5}
by successive orbifolding in the Landau-Ginzburg phase. \pano
On is familiar with mirror symmetry, but there exist more general 
perturbative target space dualities 
for $(0,2)$ models. Two at large radius completely 
different looking models can have the same Landau-Ginzburg phase.
In \cite{paper5} it was shown that also at large radius the total dimension
of the moduli spaces agrees providing evidence for the conjecture that
the models are isomorphic throughout the entire moduli space with a 
non-trivial map among complex, K\"ahler and bundle moduli. \pano 
Other important approaches to describe  $(0,2)$ models using F-theory and
equivariant sheaves are not covered here.

\section{Exactly solvable SCFTs and Distler-Kachru models}
Gepner provided exactly solvable conformal field theories (CFT) describing 
special point in the moduli space of a Calabi-Yau $(2,2)$ compactification.
In this case, up to the application of the bosonic string map 
the modular invariant partition function is left-right symmetric.
In order to find a CFT description of more general $(0,2)$ compactifications
one needs a method for  constructing really heterotic partition functions.
One way to achieve this is by using simple currents. Since two simple 
currents can be non-local to each other, the partition function
obtained after modding out these simple currents need not to be left-right 
symmetric.
As a generalisation of Gepner models we proposed the following 
CFTs
\begin{center}
\begin{tabular}{|l r r |}
\hline
             & $c$  & $\bc$  \tabroom  \\
\hline
flat space-time &2         &2   \tabroom \\
N$=$2 SCFT   &9            &9            \tabroom \\
$\left({\rm  U(1)}_2\right)^{r-3}$
             &$r-3$        &$r-3$  \tabroom \\
SO(16$-2r)\times $E$_8$
             &$16-r$       &$16-r$\tabroom \\
\hline
\end{tabular}
\end{center}
\centerline{{\bf Table 1:}~{\it Ingredients for generalised Gepner models}}
\sn
In \cite{paper2} we considered the following modular invariant 
partition function
\beq
Z\sim \vchi(\tau) M({\rm J}_{\rm GSO}) \left(\prod_i M(\Upsilon_i)\right)
       M({\rm \bJ}_{\rm GSO}) \left(\prod_{i=1}^r M({\rm J}_i)\right)
  \left(\prod_{j=1}^{r-3} M({\rm J}^j_{\rm ext})\right) \vchi(\btau),
\eeq
where the simple currents are chosen in such a way as to guarantee 
two right moving world sheet supersymmetries, one space time
supersymmetry and an extension of the gauge group from
$SO(16-2r)\times U(1)^{r-3}$ to $E_{9-r}$.
If the simple current $\Upsilon_l$ does contain factors of both NS and R type,
then the left moving supersymmetry is broken and one obtains a model
with gauge group $E_{9-r}\times E_8\times G$.  For suitable choices of the
simple current, by comparing massless spectra and chiral rings one can 
identify them as special point in the moduli space of linear sigma models. 
As an example consider the 
$(k=3)^5$ Gepner model with $r=4$ and choose
\beq
\Upsilon=\Phi^3_{0,-1}\otimes\left(\Phi^0_{0,0}\right)^4
 \otimes\Phi^{U(1)_2}_{1,2}\otimes\Phi^{SO(8)}_0,  
\eeq
having gauge group $SO(10)$ and $N_{16}=80$ generations, no antigeneration, 
$N_{10}=74$ gauge vectors and $N_1=350$ gauge singlets.
This agrees with the spectrum of the linear sigma model
\beq
 \IP_{1,1,1,1,2,2}[4,4] \leftarrow V_{1,1,1,1,1}[5],
\eeq
where the vector bundle $V$ is defined  by an exact sequence 
\beq
 0\to\ V\to\bigoplus_{a=1}^{5}{\cal O}(1)\to{\cal O}(5)\to0 .
\eeq
Generalising this example in \cite{paper3} we defined a nice subclass of 
models.
Given a Gepner model with $K_1=2\ell-1$. Let $d$ be
the lowest common multiple of the numbers 
$\lbrace K_i: i=1,\ldots,5 \rbrace$.
For models with only four factors set $K_5=0$. Then the analysis of the
chiral ring reveals that a model obtained by using the following simple
currents in the diagonal Gepner parent model
\beq
 \Upsilon=\Phi^{K_1}_{0,-1}\otimes\left(\Phi^0_{0,0}\right)^4 
  \otimes\Phi^{U(1)_2}_{1,2}\otimes\Phi^{SO(8)}_0  
\eeq
corresponds to a linear $\sigma-$model with the following data
\beq
\IP_{{2d\over 2\ell+1},{\ell d\over 2\ell+1},{d\over K_2+2},
   {d\over K_3+2},{d\over K_4+2},{d\over K_5+2}}
   \left[{\ts {(\ell+2)d\over 2\ell+1},{2\ell d\over 2\ell+1}}\right] 
   \leftarrow 
 V_{{\ts {d\over 2\ell+1},{d\over K_2+2},{d\over K_3+2},
 {d\over K_4+2},{d\over K_5+2} }}[d].
\eeq
Roughly speaking one generates $(0,2)$ data from $(2,2)$ data automatically
satisfying the non-trivial anomaly constraints. This class of models
provided a playground for further study. 
Mirror symmetry has become an important tool in exactly describing
moduli spaces of $(2,2)$ Calabi-Yau compactifications.
For $(0,2)$ models non-perturbative sigma model and target space space 
corrections
are under less control. At least in the class defined above one can generate
candidate dual pairs as for instance
\beq
 \IP_{1,1,1,1,2,2}[4,4] \leftarrow V_{1,1,1,1,1}[5] \quad\quad
 \IP_{51,60,80,65,128,128}[256,256] \leftarrow 
     V_{51,64,60,80,65}[320] .
\eeq 
Starting with a $(2,2)$ mirror pair, one applies the transformation to
get two $(0,2)$ models with still mirror symmetric spectra.  
Another approach to generate mirror pairs is by orbifolding. To this end
we developed orbifold techniques for $(0,2)$ Landau-Ginzburg models
in \cite{paper4,paper5}. We showed that by successive orbifolding of 
the model in (3) one obtains a mirror symmetric set of models. 

\section{Target space dualities}
Besides mirror symmetry, in the $(0,2)$ context one can imagine other
target space dualities at the perturbative level.
It could happen that two models defined by Calabi-Yau threefold and bundle
data $(M_1,V_1)$ and $(M_2,V_2)$ are isomorphic as superconformal
field theories. 
One way to realize such a duality exists in the framework of linear 
sigma models \cite{paper7}.
In the Landau-Ginzburg phase the superpotential reduces to
\beq
 S=\int d^2 z d\theta \left[ \Gamma^j W_j(\Phi_i) + 
                p \Lambda^a F_a(\Phi_i)\right], \quad\quad {\rm with}\
              p=\la P \ra,
\eeq   
where $W_j$ define the hypersurfaces in a weighed projective space and
$F_a$ the bundle. In our former notation this defines a model
\beq
\IP_{\omega_1,\ldots,\omega_{N_\omega}}[d_1,\ldots,d_{N_d}] 
      \leftarrow V_{n_1,\ldots,n_{N_n}}[m]. 
\eeq
The parameters $\omega_i, d_j, n_a,m$ are related
to the $U(1)$ charges of the corresponding superfields $\Phi_i,
\Gamma^j, \Lambda^a,P$ in the gauged linear sigma model. 
In (8) manifold and bundle data appear on equal footing so that 
it might be possible that two different sets of geometric data lead to
the same Landau-Ginzburg models. It was believed for some time
that the Landau-Ginzburg point is like a transition point from one
$(0,2)$ model to another \cite{paper7}.
In \cite{paper6} it was argued that a different scenario occurs, namely that
the two models are isomorphic at every point in moduli space.
The argument was based on an exact computation of the dimensions of the
geometric moduli spaces including complex, K\"ahler and bundle moduli.
As an example consider the quintic
\beq
  \IP_{4}[5]\ {\rm with\ deformation\ of}\ T 
\eeq
and a resolution of
\beq
\IP_{1,1,1,1,1,3}[4,4] \leftarrow V_{1,1,1,2}[5] .
\eeq
They have the same Landau-Ginzburg locus.
Using  techniques from toric geometry and homological algebra
one can  compute the exact dimensions of various cohomology groups.
The gauge group in both models is $E_6\times E_8$. They both have the same 
number of generations $H^1(M,V)=101$ and antigenerations $H^1(M,V^*)=1$. 
For the first model the number of complex, K\"ahler and bundle 
moduli is $H^1(M_1,T)=101$, $H^1(M_1,T^*)=1$ and $H^1(M_1,End(T))=224$
adding up to a total of 326 moduli.
For the second model the numbers are
$H^1(M_1,T)=86$, $H^1(M_1,T^*)=2$ and $H^1(M_1,End(V))=238$
amazingly adding up to 326, as well. 
In all the examples studied, the number of geometric moduli agreed completely
where of course the individual contributions of the three kinds of moduli
got exchanged. With such high dimensional moduli spaces involved it
is difficult to determine the exact map between various moduli. 
Furthermore, one might asked whether such dualities are of any use for
exact non-perturbative computations like for $(2,2)$ mirror symmetry. 
All the non-renormalization theorems holding for $(2,2)$ models are 
generically not true for $(0,2)$.

\vskip0.5cm
\noindent
{\large \bf Acknowledgements}
\smallskip
\noindent
I would like to thank Andreas Wi\ss kirchen, Rolf Schimmrigk, Sav Sethi and
Michael Flohr for their collaboration on part  of the work presented in 
this talk.

\end{document}